\documentclass[%
 reprint, 
 superscriptaddress,
 amsmath,amssymb,
 aps, pra, 
 floatfix, 
]{revtex4-2}
\usepackage{multirow}
\DeclareMathOperator\erf{erf}

\usepackage{graphicx}
\usepackage{dcolumn}

\begin{document}

\preprint{}

\title{\textbf{Interfacial defect properties of high-entropy carbides: Stacking faults, Shockley partial dislocations, and a new Evans-Polanyi-Semenov relation} 
}%

\author{Samuel E. Daigle}
\affiliation{Department of Materials Science and Engineering, North Carolina State University, Raleigh, North Carolina 27695}
\email{sedaigle@ncsu.edu}
            
\author{Stefano Curtarolo}
\affiliation{Thomas Lord Department of Mechanical Engineering and Materials Science, Duke University, Durham, North Carolina 27708}
\affiliation{Center for Extreme Materials, Duke University, Durham, North Carolina 27708}
            
\author{William G. Fahrenholtz}
\affiliation{Missouri University of Science and Technology, Rolla, Missouri 65409}
            
\author{Jon-Paul Maria}
\affiliation{Pennsylvania State University, Department of Materials Science and Engineering, State College, Pennsylvania 16802}

\author{Douglas E. Wolfe}
\affiliation{Pennsylvania State University, Applied Research Laboratory, State College, Pennsylvania 16802}
            
\author{Eva Zurek}
\affiliation{Department of Chemistry, State University of New York at Buffalo, Buffalo, New York 14260}
            
\author{Donald W. Brenner}
\affiliation{Department of Materials Science and Engineering, North Carolina State University, Raleigh, North Carolina 27695}

\date{\today}

\begin{abstract}

Using first principles calculations, $\{111\}$ intrinsic stacking fault (ISF) energies in group IVB, VB, and VIB high-entropy transition metal carbides (HETMCs) are shown to be predictable from an optimized rule of mixtures based on the atomic arrangement near the stacking fault. A composition-independent linear relationship is demonstrated between the ISF energies and the unstable stacking fault (USF) energies along the $\langle11\bar{2}\rangle\{111\}$ gamma surface slip path. This relationship represents a new application of the Evans-Polanyi-Semenov principle by treating the ISF and USF energies as analogous to the heat of reaction and transition state barrier in chemical reactions. Further, a full defect energy distribution can be obtained from the predicted ISF energies for each early transition metal HETMC. Balancing the elastic repulsion between partial dislocations with the distribution of ISF energies, we show that Shockley partial edge dislocations should remain bound for HETMCs with valence electron concentration up to 9.6, even when the average stacking fault energy is negative.

\end{abstract}

\maketitle



\section{\label{sec:intro}Background and Motivation}

The first high-entropy ceramic, a five cation rocksalt oxide, was reported in 2015~\cite{rost_entropy-stabilized_2015}. Since then, the number of high-entropy ceramics has rapidly expanded and now includes a range of different oxides, carbides, diborides, nitrides, carbo-nitrides, perovskites and other related materials~\cite{oses_high-entropy_2020, aamlid_understanding_2023, feng_high-entropy_2021, liang_vacancy_2022, jiang_new_2018, qin_new_2021, kotsonis_highentropy_2023, gurao_high-entropy_2020, zhang_review_2019, wang_processing_2022, sarkar_rare_2018, qin_high-entropy_2021, bryce_calcium_2023, sang_crystal_2023, ping_structural_2023}. One of the central questions associated with these materials is the degree to which the structural, thermal, mechanical, electronic and chemical properties can be predicted by a rule of mixtures (RoM) based on their constituent compounds, and which properties are unique to a high-entropy ceramic. For example, the rocksalt oxide with cations Mg$^{+2}$, Ni$^{+2}$, Co$^{+2}$, Cu$^{+2}$, and Zn$^{+2}$ was found to have a lattice constant that is well described by a RoM and Bader charges that are transferable between the binary, ternary, and high-entropy compositions~\cite{rak_charge_2016}. However, adding aliovalent cations such as Li$^{+1}$ and Sr$^{+3}$ results in a range of cation and oxygen charges, multiple valence states, and a reduction in thermal conductivity due to enhanced phonon scattering from the different charge states~\cite{braun_charge-induced_2018, lim_influence_2019}. For stoichiometric high-entropy carbides with transition metals (HETMCs) from groups IVB, VB, and VIB, theory predicts that lattice constants, bulk moduli, and cohesive energies are all well described by a RoM from the constitutive rocksalt binaries~\cite{lim_predicting_2023}. The carbon vacancy formation energies, on the other hand, cannot be accurately estimated from a simple RoM, but have been estimated using a neural network trained on computational energies~\cite{zhao_machine_2023}.

Of particular interest to this work is the potential for high-entropy ceramics as ultra-hard materials~\cite{sarker_high-entropy_2018, liang_achieving_2021, liang_vacancy_2022, zhang_ultra_2021, chen_stability_2019}. To that end, the gamma surface can be used as a powerful and convenient computational tool to understand slip systems in crystals which give rise to their mechanical properties. The gamma surface is generated by moving two adjacent regions of a crystal along a given interfacial plane and calculating the potential energy after relaxing the system in the direction normal to the interface. The potential energy is given as a function of the relative in-plane displacement of the slabs to produce an energy landscape equivalent to a topological map. The gamma surface was introduced by Vitek to find energetically stable stacking faults in bcc materials~\cite{vitek_intrinsic_1968}, and has since become an invaluable tool for understanding slip and related mechanical deformations. For example, intrinsic stacking fault (ISF) energies (energy valleys) are used to predict partial dislocation separations, while unstable stacking fault (USF) energies (peaks in the gamma surface) can be used to predict energetically favorable slip systems. USF and ISF energies are also used in analytic expressions to predict quantities such as fracture and twinning tendencies. Tadmor and Bernstein, for example, developed an expression for predicting twinning that includes the ratio ISF/USF~\cite{bernstein_tight-binding_2004, tadmor_first-principles_2004}.

For this work, it is important to distinguish between the gamma surface and quasi-static shear calculations where a shear deformation is applied to the entire unit cell to model the slip process. This distinction was shown for two fcc metals by Jahnatek et al.~\cite{jahnatek_shear_2009}, with the two methods producing similar USF energies, but different stresses. Hence, stress along a given slip path on the gamma surface is not necessarily equivalent to the critical stress for dislocation mobility~\cite{jahnatek_shear_2009, zhang_unstable_2023}.

Experimental and computational studies have established that the rocksalt (Fm$\bar3$m) group IVB carbides TiC, ZrC, and HfC and the group VB carbides VC, NbC, and TaC prefer to slip along  $\langle110\rangle$ directions~\cite{yu_understanding_2017}. However, the group IVB carbides slip along \{110\} planes, while the group VB carbides slip along \{111\} planes. Studies of mixed IVB-VB carbides have reported either slip on both planes, or on \{111\} planes exclusively~\cite{smith_phase_2018, csanadi_hardness_2021}. By examining the gamma surface with energies calculated from Density Functional Theory (DFT), Thompson, Weinberger and co-workers have shown that the transition from slip on \{110\} to \{111\} planes can be attributed to a lowering of the barrier along the $\langle112\rangle$\{111\} slip direction due to a decrease in the high-symmetry ISF energy as composition goes across the columns~\cite{yu_understanding_2017, de_leon_bonding_2015}. Further, slip along the [11$\bar2$] direction to the low energy ISF followed by slip along [$\bar12\bar1$] on \{111\} planes leads to a net slip along the [$01\bar1$] direction. The decomposition of $\langle110\rangle$\{111\} slip is illustrated for fcc Ni by Shang et al. using an alias shear deformation model~\cite{shang_temperature-dependent_2012}.

Ma et al. have used DFT to calculate the energies along the slip and twinning paths on \{111\} planes for rocksalt HETMCs containing Zr, Nb, Ta, Hf, Ti, and V~\cite{ma_effect_2021}. They report that USF energies for compositions with mixed cations are lower than those for the single-metal carbides, suggesting that dislocation nucleation would be more prevalent in the latter compared to the binaries. Our calculations in this work do not reproduce this result. Instead, we find that the binaries and the high-entropy compositions studied all fall along well-behaved RoM relations for ISF and USF energies. This discrepancy likely stems from an inadequate sampling of the configuration space in the prior work, as well as reported USF energies for the single-metal carbides which are higher than those reported in literature~\cite{yu_ab-initio_2016} by a factor of 2. In addition, the Ma et al. study did not characterize the shift in the position of the USF away from the midpoint between the pristine crystal and the ISF configuration. As shown below, accounting for this shift is critical to accurately calculating the USF energy, and leads to a previously unreported relationship between the ISF and USF. 

The intent of this research is to explore the degree to which ISF and USF energies along the $\langle11\bar2\rangle$\{111\} gamma surface path in stoichiometric group IVB-VB-VIB HETMCs can be estimated from a RoM using the corresponding values for the single-metal rocksalt carbides. The HETMCs compositions in this study are listed in Table~\ref{tab:comps}.

\begin{table}[!htpb]
\centering
\caption{\label{tab:comps} High-entropy carbide compositions studied for each calculation. Compositions are experimentally single-phase~\cite{kaufmann_discovery_2020,sarker_high-entropy_2018,divilov_disordered_2024} except as noted.}
    \begin{ruledtabular}
    \begin{tabular}{cll}
        \multicolumn{1}{c}{Calculation} & \multicolumn{2}{c}{Compositions}\\
        \hline
        \multirow{7}{*}{ISF} & (Cr,Mo,Nb,Ta,W)C & (Cr,Mo,Nb,V,W)C \\
         & (Hf,Mo,Nb,Ta,W)C\footnote[1]{Phase data not available.} & (Hf,Mo,Ta,W,Zr)C\footnote[2]{Multi-phase~\cite{sarker_high-entropy_2018}.} \\
         & (Hf,Mo,Ti,W,Zr)C & (Hf,Mo,V,W,Zr)C\footnotemark[2] \\
         & (Hf,Nb,Ta,Ti,V)C & (Hf,Nb,Ta,Ti,W)C \\
         & (Hf,Nb,Ta,Ti,Zr)C & (Hf,Ta,Ti,W,Zr)C \\
         & (Mo,Nb,Ta,Ti,V)C & (Mo,Nb,Ta,V,W)C \\ 
         & (Mo,Nb,Ta,W,Zr)C\footnotemark[1] & (Nb,Ta,Ti,V,W)C \\ \hline
        $\langle11\bar2\rangle\{111\}$ & \multirow{2}{*}{(Hf,Nb,Ta,Ti,Zr)C} & \multirow{2}{*}{(Mo,Nb,Ta,V,W)C} \\
        slip path & & \\ \hline
        \multirow{2}{*}{USF} & (Hf,Nb,Ta,Ti,V)C & (Hf,Nb,Ta,Ti,Zr)C \\
        & (Mo,Nb,Ta,V,W)C & (Nb,Ta,Ti,V,W)C \\
    \end{tabular}
    \end{ruledtabular}
\end{table}

As discussed in more detail below, three general conclusions come from the research. First, based on the DFT results, a RoM with an optimized weighting scheme for atomic layers near the stacking fault provides energies within $\pm 0.2$ J/m$^2$ for both the average and local ISF energies. Second, by including group VIB elements, the average ISF energy of a HETMC can be negative even in compositions which are predicted by entropy descriptors and confirmed by experiment to form stable single-phase materials~\cite{kaufmann_discovery_2020, divilov_disordered_2024}. A random cation arrangement can lead to regions of positive and negative stacking fault energies, which we evaluate statistically. Similar to other work on lattice roughness and slip in high-entropy materials~\cite{utt_origin_2022, wang_enhanced_2020, shih_stacking_2021}, it is expected that the upper tails of these energy distributions contribute to the Peierls stress that prevents separation of partial dislocations into extended stacking faults. Finally, it was found that the USF does not occur at the high-symmetry midpoint between the ideal system and the ISF, but rather is shifted along the slip path. This peak shift is apparent from literature gamma surfaces for transition metal carbides, such as those presented by Yu et al.~\cite{yu_understanding_2017}, but has not previously been quantified. We find that the peak shift can be defined as a monotonic function of the ISF, enabling efficient calculation of the USF energy and uncovering a linear relationship between USF and ISF energies spanning net negative and net positive ISF values. 

As a result of these relationships, the ISF energy can be found via a RoM, while the associated USF and the peak shift can be estimated as a function of the ISF. The linear relation between the USF and ISF is analogous to the well-established Evans-Polanyi-Semenov (EPS) relation~\cite{evans_inertia_1938}, where for a class of similar chemical reactions, the energy barrier is linearly proportional to the heat of reaction. 

\section{\label{sec:calc}Calculation Details}

DFT energies were calculated using the plane-wave projector-augmented-wave pseudopotential methods~\cite{blochl1994paw} in the Vienna Ab initio Simulation Package~\cite{kresse1993ab-initio, kresse1996efficiency, kresse1996efficient}. The exchange-correlation potential used was the Perdew et al.~\cite{perdew1996gga} parameterization of the generalized gradient approximation. The plane-wave basis cutoff energy was 520 eV with a 6×6×1 $\Gamma$-centered k-point mesh. Spin polarization was enabled for all simulations to account for possible magnetism in the group VIB transition metals.

Gamma surface configurations for the rocksalt carbides were constructed as translations of an 80-atom slab with two (111) free surfaces separated by at least 12 Å of vacuum. The $\mathbf{a}$ and $\mathbf{b}$ supercell vectors were defined as $\frac{1}{2}[11\bar{2}]$ and $[\bar110]$, respectively. For each composition, 50 random arrangements of the cation sites were generated. Relaxations were performed with respect to the atomic positions in the [111] direction, while keeping the positions fixed in $\mathbf{a}$ and $\mathbf{b}$. Lattice parameters for the high-entropy carbide compositions were determined by the average result of energy minimization performed on 10 randomly arranged 80-atom bulk rocksalt structures with supercell vectors [210], $[\bar{1}20]$, and [002]. 

All calculations in these studies were performed with full carbon stoichiometry. Transition metal carbides are often carbon deficient, especially at elevated temperatures, and carbon vacancies can affect stacking fault energies~\cite{ding_influence_2014}. However, vacancy effects in small supercells require additional considerations and are not captured in this work.

\begin{figure*}[!t]
\centering
\includegraphics[width=.9\textwidth]{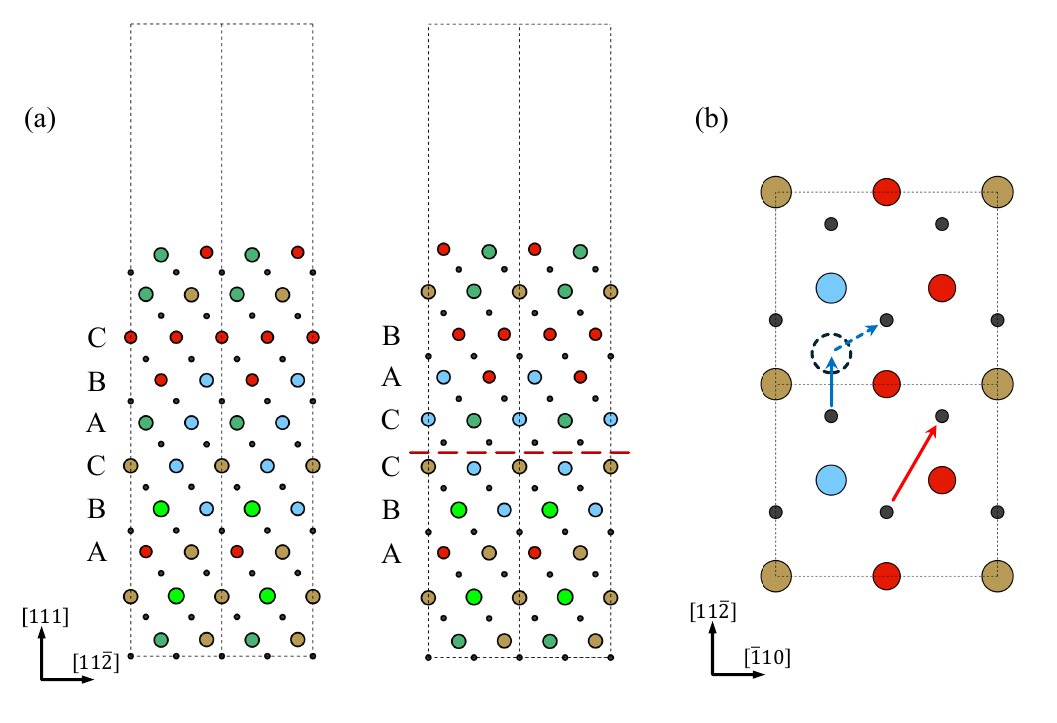}
\caption{\label{fig:supercell} DFT supercell structures for gamma surface calculations. The cation stacking sequence of the pristine and ISF configurations are shown in (a) with the stacking fault indicated by the red dashed line. The slip directions of the carbon plane sliding over the cation plane are indicated in (b) for $\langle11\bar2\rangle$ (blue arrow) and $\langle110\rangle$ (red arrow) directions. The solid and dashed blue arrows illustrate the decomposition of $\langle110\rangle\{111\}$ slip in fcc materials.} 
\end{figure*}

Gamma surface energies $\gamma$ were calculated for each configuration and the constituent binaries using the expression
\begin{eqnarray}
\gamma = \frac{E_\gamma-E_{slab}}{A}\,
\end{eqnarray}
where $E_\gamma$ and $E_{slab}$ are the energy of the translated and vacuum slab configurations, respectively, and $A$ is the cross-sectional area of the supercell in the (111) plane. The local environment of any point on the (111) gamma surface can be defined by the translation vector between the upper and lower portions of the slab and the local atomic environment in the nearby (111) cation planes. An example supercell used for the slab and ISF calculations is shown in Fig.~\ref{fig:supercell}.

\section{\label{sec:results}Results and Discussion}

The compositions in Table~\ref{tab:comps} were studied to determine the relationship between ISF energy, USF energy, and the USF position along the $\langle11\bar2\rangle\{111\}$ slip path. Because the shape of the gamma surface varies with composition at the stacking fault, a larger set of calculations was required to find the USF peak position for a given configuration. Compositions were divided into three simulation regimes to efficiently characterize the critical points of the slip path.

\subsection{\label{sec:sfe}Stacking Fault Energies}

Stacking faults in high-entropy compounds differ from those of single-component materials in that the energy is dependent on the arrangement of nearby atoms in the structure. Similar to vacancies and surfaces in high-entropy alloys and ceramics, the defect energy can be defined as a distribution of energies based on the local atomic environment~\cite{lim_predicting_2023, zhao_machine_2023, riley_vacancy-driven_2023, daigle_statistical_2020, kristoffersen_local_2022}, in this case defined as the composition along the interface. To capture the energy dependence on the local arrangement of atoms, a sample of atomic configurations is required for each composition. The sample can then be used to fit a model and predict the total energy distribution across the configurational space, enabling analysis of the impact of defects on bulk properties and phenomena~\cite{daigle_statistical_2020}. Here, DFT energies were calculated for 50 random ISF configurations of 14 HEC compositions (see Table~\ref{tab:comps}).

For these HEC compositions, the ISF energy is found to be approximated by a weighted RoM of the single-metal carbide stacking fault energies for atoms within 0.5~nm of the defect plane as
\begin{eqnarray}
\gamma_\omega^{isf} = \sum_i\sum_k\frac{\omega_k n_{ik} \gamma_i^{isf}}{n_k} .
\label{eqn:omega}
\end{eqnarray}

Here, $n_{ik}$ are the number of atoms of each alloying element $i$ in each layer $k$, $\gamma_i^{isf}$ are the \{111\} stacking fault energies in the single-metal carbides, and $\omega_k$ are the fitted layer weights. For the 80-atom supercell in this work, $n_k=4$, and the normalized weights $\omega_{1-4}$ are found to be 0.589, 0.321, 0.045, and 0.045, respectively. While the layer weights are calculated as fitted parameters, they decrease with increasing distance from the defect, corresponding to a qualitative decay function representing some unknown physical interaction. The energies of the single-metal carbide stacking faults are given in Table~\ref{tab:binaries}.

\begin{table}[!t]

\centering
\caption{\label{tab:binaries} Single-metal carbide stacking fault energies $\gamma^{isf}$ (J/m$^2$) used in Eq.~\ref{eqn:omega}.}
    \begin{ruledtabular}
    \begin{tabular}{llllll}
        \multicolumn{2}{c}{IVB} & \multicolumn{2}{c}{VB} & \multicolumn{2}{c}{VIB} \\
        \hline
        Ti & 2.757 & V & 0.453 & Cr & -0.777 \\
        Zr & 2.645 & Nb & 0.667 & Mo & -1.002 \\
        Hf & 2.691 & Ta & 0.572 & W & -1.353 \\
    \end{tabular}
    \end{ruledtabular}
\end{table}

Here we note that the ISF energies for group VIB metal carbides are all negative, as these compositions prefer the hexagonal WC or Mo$_2$C structure (P$\bar6$m2) to the rocksalt structure (Fm$\bar3$m). In HETMCs, these elements can be present as equimolar components~\cite{sarker_high-entropy_2018,kaufmann_discovery_2020}, resulting in negative stacking faults for certain configurations within the stable structure. The consequences of negative stacking faults in HETMCs are considered in Section~\ref{sec:shockly}.

\begin{figure}[!t]
\centering
\includegraphics[width=.45\textwidth]{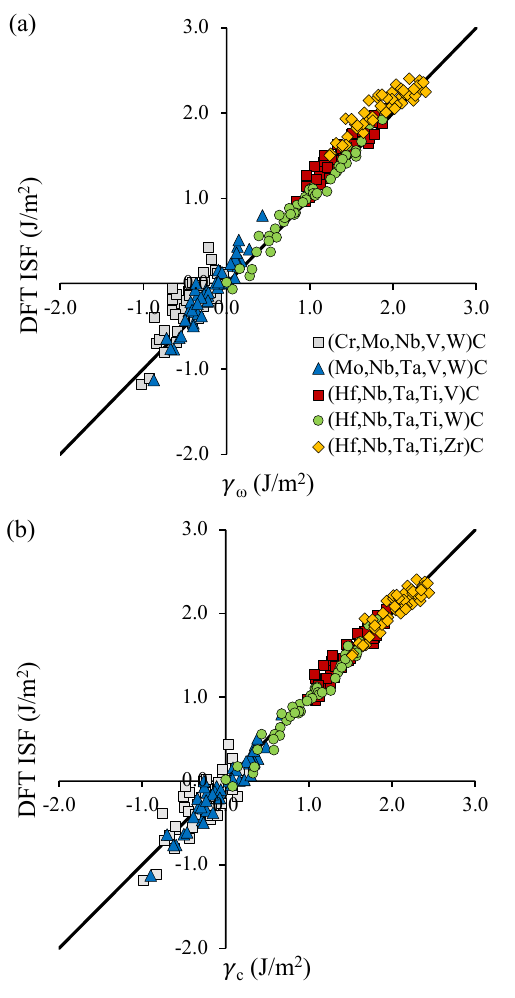}
\caption{\label{fig:fit} DFT calculated ISF energies as a function of RoM estimates fit to (a)~layer weights only using Eq.~\ref{eqn:omega} and (b)~weighted RoM with empirical corrections using Eq.~\ref{eqn:isf}.}
\end{figure}

Using the single-metal carbide stacking fault energies and the given weights $\omega_k$, the optimized RoM gives a reasonable approximation of the high-entropy ISF energies, as plotted for five compositions in Fig.~\ref{fig:fit}a. However, comparing the RoM prediction to the DFT data in Fig.~\ref{fig:fit}, it is apparent that compositions with high and low ISF energies are underestimated by a RoM and that the distribution of energies within each composition is itself biased. This bias is accounted for by fitting two quadratic corrections, first to the bulk RoM ISF energies $\bar{\gamma}$ for each composition and second to the local weighted RoM $\gamma_\omega$ as expressed in Eq.~\ref{eqn:omega}. The results are plotted for the same five compositions in Fig.~\ref{fig:fit}b. These empirical corrections reduce the root mean squared error by 26\% from 182 mJ/m$^2$ to 135 mJ/m$^2$. The error for high [(Hf,Nb,Ta,Ti,Zr)C] and low [(Cr,Mo,Nb,V,W)C] valence electron concentration (VEC) compositions are reduced by 55\% and 27\% respectively. The fitted model is given as
\begin{eqnarray}
\gamma_{c}^{isf} &=(&0.456\bar{\gamma}-0.335)^2 \nonumber \\
& &  -(0.303\gamma_\omega-1.835)^2+3.393,
\label{eqn:isf}
\end{eqnarray}
with model errors for each composition presented in Table~\ref{tab:fit}. In all cases, the ISF energies for stacking fault configurations in HETMCs are well described as an average of the single-metal carbide constituents near the interface. 

These results compare favorably to predictions from axial nearest-neighbor Ising (ANNI) and axial next-nearest-neighbor Ising (ANNNI) models, which approximate the ISF as the energy associated with a combination of fcc, hcp, and dhcp planes corresponding to the local stacking sequence.~\cite{breidi_first-principles_2024,denteneer_stacking-fault_1987} Accounting for the layer interactions with the carbon sublattice, these models are given as
\begin{eqnarray}
\gamma_{ANNI}^{isf} = \frac{4(E_{hcp}-E_{fcc})}{a^2\sqrt{3}} 
\label{eqn:anni}
\end{eqnarray}
and 
\begin{eqnarray}
\gamma_{ANNNI}^{isf} = \frac{2(E_{hcp}+2E_{dhcp}-3E_{fcc})}{a^2\sqrt{3}} ,
\label{eqn:annni}
\end{eqnarray}
where $a$ is the fcc lattice constant. These equations can be modified for the high-entropy case by summing over the atoms occupying each layer of the stacking fault structure, with $E_{hcp}$ corresponding to the weighted RoM plane $\omega_1$ and $E_{dhcp}$ corresponding to $\omega_2$ and $\omega_3$.

The Ising models have proved useful for stacking fault energy estimates in metallic systems,~\cite{breidi_first-principles_2024} however their accuracy for the early transition metal carbides is inconsistent. Specifically, the ANNI model overestimates the ISF energy for the group VB carbides while the ANNNI model underestimates the ISF for the group IVB carbides. The HETMC and single-metal carbide Ising model errors are included in Table~\ref{tab:fit}.

\begin{table}[!t]
    \caption{\label{tab:fit} Root mean squared errors (J/m$^2$) for HETMC ISF energies from weighted RoM models $\gamma_\omega^{isf}$ and $\gamma_c^{isf}$, and from near-neighbor Ising models $\gamma_{ANNI}^{isf}$ and $\gamma_{ANNNI}^{isf}$.}
    \centering
    \begin{ruledtabular}
    \begin{tabular}{lcccc}
    Composition & $\gamma_\omega^{isf}$ & $\gamma_c^{isf}$ & $\gamma_{ANNI}^{isf}$ & $\gamma_{ANNNI}^{isf}$ \\
    \hline
    (Cr,Mo,Nb,Ta,W)C & 0.294 & 0.220  & 0.285 & 0.207\\
    (Cr,Mo,Nb,V,W)C & 0.263 & 0.192 & 0.286 & 0.234 \\
    (Hf,Mo,Nb,Ta,W)C & 0.164 & 0.121 & 0.269 & 0.247 \\
    (Hf,Mo,Ta,W,Zr)C & 0.163 & 0.135 & 0.265 & 0.261 \\
    (Hf,Mo,Ti,W,Zr)C & 0.111 & 0.124 & 0.268 & 0.281 \\
    (Hf,Mo,V,W,Zr)C & 0.146 & 0.126 & 0.291 & 0.304 \\
    (Hf,Nb,Ta,Ti,V)C & 0.171 & 0.103 & 0.450 & 0.176 \\
    (Hf,Nb,Ta,Ti,W)C & 0.082 & 0.088 & 0.331 & 0.279 \\
    (Hf,Nb,Ta,Ti,Zr)C & 0.227 & 0.102 & 0.188 & 0.176 \\
    (Hf,Ta,Ti,W,Zr)C & 0.145 & 0.117 & 0.226 & 0.251 \\
    (Mo,Nb,Ta,Ti,V)C & 0.146 & 0.151 & 0.547 & 0.314 \\
    (Mo,Nb,Ta,V,W)C & 0.163 & 0.123 & 0.388 & 0.209 \\
    (Mo,Nb,Ta,W,Zr)C & 0.170 & 0.113 & 0.308 & 0.265 \\
    (Nb,Ta,Ti,V,W)C & 0.137 & 0.136 & 0.489 & 0.289 \\
    CrC &  &  & -0.036 & 0.011 \\
    HfC &  &  & -0.043 & -0.405 \\
    MoC &  &  & -0.122 & 0.114 \\
    NbC &  &  & 0.890 & 0.380 \\
    TaC &  &  & 0.907 & 0.345 \\
    TiC &  &  & -0.179 & -0.586 \\
    VC &  &  & 0.996 & 0.283 \\
    WC &  &  & -0.278 & 0.066 \\
    ZrC &  &  & -0.209 & -0.455 \\
    \end{tabular}
    \end{ruledtabular}
\end{table} 

\begin{figure}[!t]
\centering
\includegraphics[width=.45\textwidth]{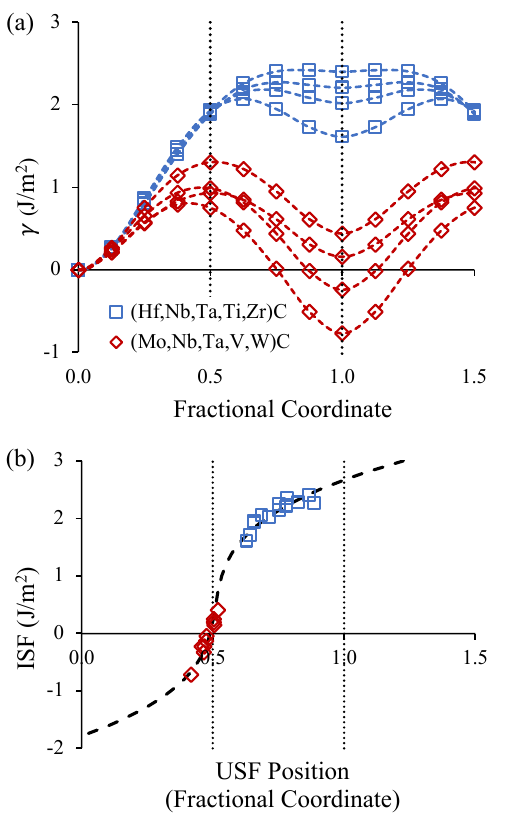}
\caption{\label{fig:slip} USF peak shift shown as (a) gamma surface energies along the $\frac{1}{6}\langle11\bar2\rangle$\{111\} slip direction, and (b) interpolated USF positions with the initial polynomial \mbox{N-R} estimate. Dashed vertical lines indicate the midpoint and ISF positions.} 
\end{figure}

The USF energy is another useful tool for understanding dislocation slip in fcc materials. However, because it is not necessarily at a high-symmetry position, it can be expensive to calculate in DFT compared to the ISF. This is illustrated by the energies along the $\frac{1}{6}\langle11\bar2\rangle$\{111\} slip path for four arrangements of two high-entropy compositions plotted in Fig.~\ref{fig:slip}. 

While the intrinsic stacking fault is always at the high-symmetry position, the USF can vary significantly from the center of the slip path depending on the composition and the cation arrangement. Using a traditional approach of scanning the gamma surface along the dislocation vector can require dozens of calculations to find the accurate maximum value for the energy barrier. This becomes especially expensive for high-entropy materials, where multiple configurations must be considered.

Here, an estimate of the USF peak location is combined with a Newton-Raphson (N-R) root-finding algorithm to predict the coordinates of the barrier on the gamma surface and minimize the computation time required to simulate the relevant configurations. Using a N-R method, the root of a differentiable function can be found with increasing accuracy by calculating the x-intercept of the tangent line at an initial guess and iterating until sufficient accuracy has been obtained. In the case of the USF, we must consider the second derivative of the gamma surface to locate the local maximum of the function.

The shape of the energy surface is characterized along the $\langle11\bar2\rangle$ vector by evaluating a coarse mesh of the gamma surface for ten random configurations of the two compositions (Hf,Nb,Ta,Ti,Zr)C and (Mo,Nb,Ta,V,W)C, representing the range of ISF energies for the HETMCs (Fig.~\ref{fig:slip}). The USF is near the high-symmetry position at $\frac{1}{12}[11\bar2]$ for (Mo,Nb,Ta,V,W)C, but is shifted toward the ISF for (Hf,Nb,Ta,Ti,Zr)C. From these two compositions, a preliminary estimate for peak shift is obtained as a third order polynomial of interpolated peak position with respect to the ISF. This function is used to initiate a N-R search wherein the DFT energy is calculated at the predicted peak and two adjacent positions to find the finite differences second derivative of energy with respect to position. Extrapolating this slope to the x-intercept gives the improved estimate for the USF peak location, where the final DFT energy can be obtained. Because the first derivative of the gamma surface is nearly linear in the region surrounding the USF, a single N-R iteration is sufficient. 

Plotted in Fig.~\ref{fig:eps}
are the calculated USF energies on the $\langle11\bar2\rangle\{111\}$ gamma surface as a function of ISF energy. For the data plotted in Fig.~\ref{fig:eps}a, the energy is calculated at the midpoint along the slip direction ($\frac{1}{12}[11\bar2]$), while the USF energy plotted in Fig.~\ref{fig:eps}b is the \mbox{N-R} estimated maximum along the slip path. This data illustrates the importance of USF peak shift when calculating the energy barrier and uncovers a linear relationship between ISF and USF energies that is discussed in the following section in terms of the EPS principle.

The tested HETMC compositions are taken as a representative sample of the 126 quinary high-entropy carbides containing the group IVB, VB, and VIB transition metals. Applying the observed relationships to these materials, the full ISF and USF energy distributions, derived in Section~\ref{sec:shockly}, can be predicted to give insights into a wide range of compositions. For compositions with more or fewer metallic components, differences in the lattice strain may result in small deviations from the empirical fitting parameters calculated in this work.

\subsection{\label{sec:eps}Evans-Polanyi-Semenov Relations}

The EPS relation (also called the Bell–Evans–Polanyi or Brønsted–Evans–Polanyi principle) states that for a set of related chemical reactions, the activation energy $E_A$ for a chemical process is proportional to the heat of reaction $\Delta H$ as
\begin{eqnarray}
    E_A = E_A^0 + \alpha\Delta H\,
\end{eqnarray}
where  $E_A^0$ and $\alpha$ are constants~\cite{evans_inertia_1938}. This simple relation has impacted multiple areas of chemistry, including combustion~\cite{meng_theoretical_2020}, organic synthesis~\cite{pla_when_2020, wubbels_bellevanspolanyi_2015}, photo-chemistry~\cite{chen_broadening_2021, liu_correlation_2021}, polymerization~\cite{sheka_virtual_2023}, interstellar chemistry~\cite{garrod_formation_2022}, heterogeneous catalysis~\cite{cheng_bronstedevanspolanyi_2008, agarwal_coverage-dependent_2022, pascucci_understanding_2019, kropp_bronstedevanspolanyi_2019}, gel degradation~\cite{jia_thermal_2021}, surface diffution~\cite{phuthi_accurate_2023}, energetic materials sensitivity~\cite{zeman_modified_2002, luo_ignition_2019, cawkwell_understanding_2022}, materials discovery~\cite{zurek_discovering_2016}, and fundamental chemical kinetics~\cite{barrales-martinez_deeper_2022, roy_bell-evans-polanyi_2008, shestakov_quantum-chemical_2003}. Adding to this body of work, we propose the ISF and USF energies as an analogous set of quantities to the heat of reaction and activation energy, respectively. 

One central premise of HETMCs has been that their properties are continuously tunable by a RoM through the IVB-VIB elements. As an example, VEC is a quantity calculated as a RoM from constituent elements, used both as a theoretical construct within the electronic band structure~\cite{jhi_electronic_1999} and as a way of accounting for composition, including cations as well as carbon and nitrogen fraction~\cite{balasubramanian_valence_2018}. In the context of the EPS relation for stacking fault energies, it is relevant to explore (1) whether the EPS relation is maintained for the continuous range of ISF and USF values on a gamma surface, and if so (2) whether the $E_A^0$ and $\alpha$ values are constant or composition dependent. 

\begin{figure}[!t]
\centering
\includegraphics[width=.45\textwidth]{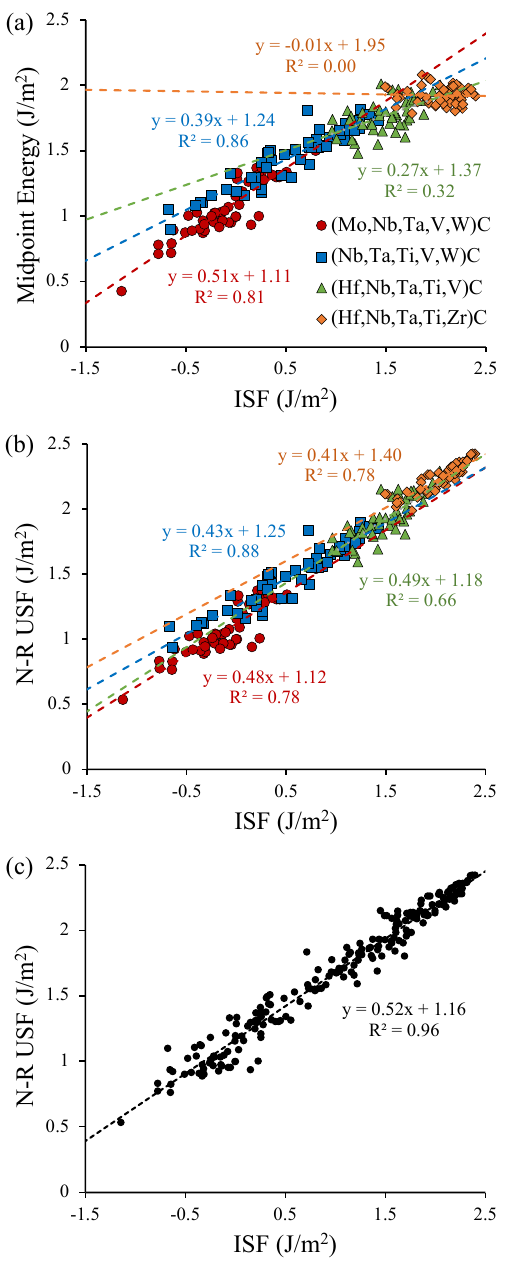}
\caption{\label{fig:eps} High-entropy carbide USF energies as a function of ISF. Energies are calculated at (a) the midpoint along the slip path, and (b and c) the N-R predicted USF position.}
\end{figure}

The energy at the high-symmetry position, plotted in Fig.~\ref{fig:eps}a, is grouped by composition with a separate linear fit for each. For (Mo,Nb,Ta,V,W)C and (Nb,Ta,Ti,V,W)C, the data is relatively well described by an EPS relation, but with a composition-dependent slope $\alpha$ that approaches zero with increasing ISF energy. As noted, however, the energy at this position does not represent the gamma surface energy barrier. After accounting for the peak shift, the energy at the \mbox{N-R} predicted USF position (Fig.~\ref{fig:eps}b) exhibits a linear EPS relation for all compositions with similar $\alpha$ and intercept $E^0_A$ values. In Fig.~\ref{fig:eps}c the combined data is plotted for all compositions at the maximum energy estimate along the slip path, indicating a composition-independent EPS relation between the ISF and USF energies when the USF peak shift is properly taken into consideration.

\begin{figure}[!t]
\centering
\includegraphics[width=.45\textwidth]{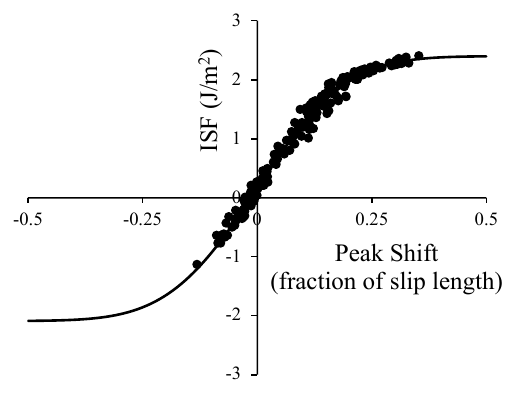}
\caption{\label{fig:erf} USF peak shift from \mbox{N-R} search along $\langle11\bar2\rangle\{111\}$ gamma surface path. Fitting function obtained from Eq.~\ref{eqn:erf} using data from the four compositions specified in Table~\ref{tab:comps}.}
\end{figure}

Plotted in Fig.~\ref{fig:erf} are the distances between the USF position estimates and the midpoint along the $\langle11\bar2\rangle\{111\}$ gamma surface path for the data in Fig.~\ref{fig:eps}c. There is a monotonic, continuous relationship between the position of the USF (the transition state) and the ISF energy (the reaction product). For increasing ISF, the USF occurs later along the slip direction, asymptotically approaching the ISF position as the ISF and USF energies converge. This behavior is approximated by the error function plotted along the length of the slip path, where
\begin{eqnarray}
    I(x) = B\erf(\alpha x) + \Delta
\label{eqn:erf}
\end{eqnarray}
for fitting parameters $B=2.245$ J/m$^2$, $a = 4.605$, and $\Delta = 0.559$ J/m$^2$. Although the physical interpretation of an error function is unclear, it provides a good quantitative fit to the data (a logistic function provides a similar fit). In particular, the asymptotic behavior of the error function is consistent with the range of possible positions of the USF along the slip path, existing only on the region between the ideal structure and the ISF. 

This function enables the evaluation of critical points on the gamma surface while minimizing the number of necessary calculations along the slip path. Given an ISF energy, which is well approximated by a weighted RoM, one can identify the position of the USF and perform a single DFT calculation without the need for an iterative search algorithm along the slip path.

\begin{figure}[!t]
\centering
\includegraphics[width=.45\textwidth, trim=0 0 -5mm 0]{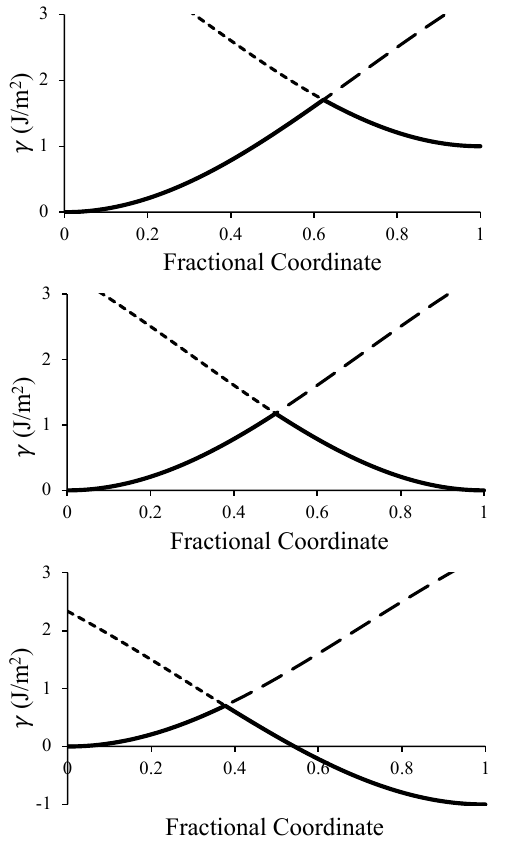}
\caption{\label{fig:overlap-model} Shifted Gaussian energy well model along the $\frac{1}{6}\langle11\bar2\rangle$\{111\} slip direction for (top)~positive, (middle)~zero, and (bottom)~negative ISF.}
\end{figure}

The relationships in Fig.~\ref{fig:eps} and Fig.~\ref{fig:erf} can be conceptualized as overlapping energy wells~\cite{evans_inertia_1938, denisov_new_1997}, illustrated in Fig.~\ref{fig:overlap-model}. In this framework, two energy wells of equal depth correspond to local energy minima on the gamma surface along the $\langle11\bar2\rangle\{111\}$ sliding direction. One curve has a minimum at the ideal structure, while the other has a minimum at the ISF and is shifted uniformly in energy according to the ISF energy. The energy along the path within this model is taken as the minimum of the two curves at any given point. Fig.~\ref{fig:overlap-model} depicts the model for the three cases of ISF$>$0, ISF$=$0, and ISF$<$0, with the barrier (USF) at the intersection of the two curves (the maximum in the solid curve). The energy barrier and its position along the slip direction both increase with higher ISF energy, consistent with the simulation data.

\begin{figure}[!t]
\centering
\includegraphics[width=.45\textwidth]{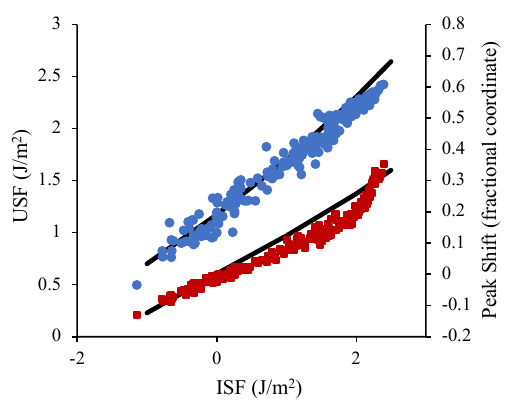}
\caption{\label{fig:overlap-fit} Maximum USF energies (blue circles) and corresponding peak shift along the $\frac{1}{6}\langle11\bar2\rangle\{111\}$ gamma surface path (red squares). Solid lines are the two-parameter fit for a simple energy curve overlap model using Eq.~\ref{eqn:gauss}.}
\end{figure}

The energy wells in the analysis for Fig.~\ref{fig:overlap-fit} are calculated as a two-parameter fit assuming identical Gaussian functions of the form
\begin{eqnarray}
    E(x) = \frac{A}{\sigma\sqrt{2\pi}}\left[1-\exp\left[\frac{1}{2}\left(\frac{x-x_0}{\sigma}\right)\right]\right]
\label{eqn:gauss}
\end{eqnarray}
for the overlapping curves. Here $x_0=0$ for the curve starting at the ideal structure, $x_0=1$ for the curve starting at the ISF, and the energy well depth is defined by $\sigma=0.7$ and $A=9.15$ J/m$^2$. This simple approximation is able to qualitatively describe the relationships discussed in this work, both between ISF and USF energies, and between ISF energy and the shift in USF position along the $\langle11\bar2\rangle\{111\}$ direction. 

Similar fits to the data can be achieved with other potential energy approximations such as quadratic energy wells, sine fitting, or a Lennard-Jones potential. None of the approximations tested, however, offer a clear advantage over the two-parameter Guassian fit.

\subsection{\label{sec:shockly}Shockley Partial Dislocation Separations}

\begin{figure*}[!t]
\centering
\includegraphics[width=.85\textwidth]{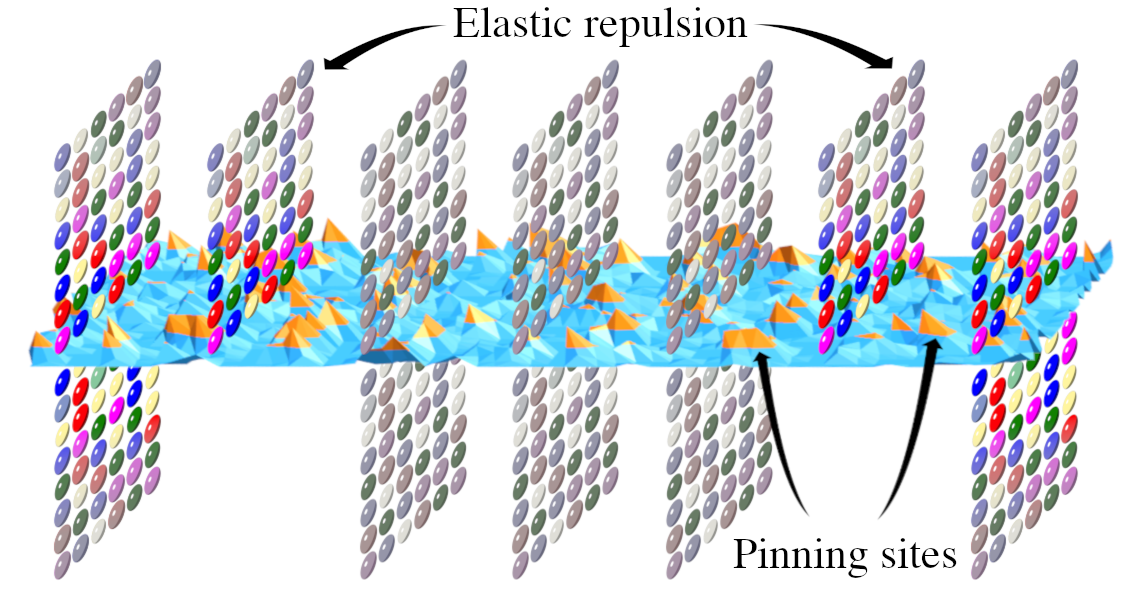}
\caption{\label{fig:partials} Depiction of two Shockley partial dislocations separated by a stacking fault along the horizontal plane. Brightly colored atoms at the edge dislocations correspond to the local atomic environment that contributes to the Peierls stress, with blue regions representing negative local stacking fault energies and orange peaks on the energy landscape representing the high energy pinning sites preventing partial dislocation separation.}
\end{figure*}

Pairs of Shockley partial dislocations in an fcc crystal are separated by a stacking fault at an equilibrium distance determined from a balance between the elastic repulsion of the dislocation cores and the stacking fault energy. One of the interesting phenomena in high-entropy materials is that stable single-phase compositions can exhibit negative stacking fault energies~\cite{shih_stacking_2021, pei_decoupling_2021}. In non-random crystals, negative stacking fault energies would suggest that Shockley partial dislocations should move as far apart as possible, because both the elastic repulsion and the stacking fault formation energy would favor infinite distances. However, large concentrations of extended stacking faults are not always observed experimentally. As an example in a HETMC, the simulations in Section~\ref{sec:sfe} indicate that (Mo,Nb,Ta,V,W)C has an average stacking fault energy of -0.09 J/m$^2$ such that the elastic repulsion between any Shockley partials should drive the formation of abundant stacking faults. Experimentally, however, this composition exhibits sharp x-ray diffraction peaks that suggest a low defect density~\cite{sarker_high-entropy_2018}. 

One suggestion to account for the apparent lack of stacking faults associated with Shockley partials is that ``roughness'' from the atomic disorder in high-entropy materials inhibits the ability of the partials to fully separate~\cite{utt_origin_2022, wang_enhanced_2020}. While the relation between Peierls stress and stacking fault energy is unclear~\cite{pei_decoupling_2021}, it is reasonable to assume that this roughness is related to the variations in stacking fault energies from the fluctuations in composition along the interface as the partial dislocations attempt to separate. 

This concept is illustrated in Fig.~\ref{fig:partials}, which shows a schematic of two Shockley partial edge dislocations separated by a stacking fault along the blue plane. The peaks on the energy landscape represent a measure of the stacking fault energy contributions to the Peierls stress. A negative stacking fault energy should lower the Peierls stress, while a positive stacking fault energy (denoted by the orange peaks) should increase the Peierls stress. Because the extra half plane exists along the length of an edge dislocation, the positive regions will pin the plane at different points despite the negative stacking fault regions exhibiting a lower local Peierls barrier. Taken to an extreme, even when the net stacking fault energy is negative, positive regions may prevent the Shockley dislocations from moving. 

In a set of atomic simulations by Shih et al.~\cite{shih_stacking_2021} modeling random NiCo alloys with dissociated Shockley partial edge dislocations and negative stacking faults, the partial separation distances remained finite at low temperatures. For all compositions tested, the average equilibrium separation distance was less than the distance at which the mean plus one standard deviation ($\mu+\sigma$) of the stacking fault energy distribution balanced the repulsive elastic stress. The mechanism for the stabilization of negative stacking faults, however, is not well-established. Shih et al. propose a force balance with an increased contribution from the solute atoms acting on the dislocation cores~\cite{shih_stacking_2021}, while Ding et al. find that the stacking fault energy in NiCoCr is dominated by chemical short-range order (SRO)~\cite{ding_tunable_2018}. Baruffi et al. report a contradictory result, finding that SRO and solute pinning effects would be insufficient to stabilize negative stacking faults in a model NiCoCr alloy~\cite{baruffi_equilibrium_2023}. Despite disagreement on the mechanism driving this behavior, the literature agrees that the effective stacking fault energy in medium- and high-entropy compositions significantly exceeds the average of the ISF distribution. For the HETMC stacking fault analysis in this work, we take the result from Shih et al. and set a lower bound estimate for energy associated with partial dislocation separation as the mean plus one standard deviation of the ISF distribution.

For this ``mean plus one'' model, we generalize the trend for partial edge dislocation separation length $s$ with the expression
\begin{eqnarray}
    s<\frac{Gb^2}{2\pi(\mu+\sigma)}\,
    \label{eqn:partial}
\end{eqnarray}
where $G$ is the shear modulus, $b$ is the Burger’s vector. 

Sampling from a random distribution of interface cation arrangements, a predicted distribution of ISF energies can be generated for each composition. Critically, the distribution is dependent on the volume that is considered to be part of the local environment. Taken to both extremes for the five cation system, the distribution of energies for a single-atom volume will consist of five sharp peaks corresponding to the five constituent elements, each accounting for 20\% of the total distribution. Similarly, as the volume of the local environment increases to infinity, the distribution will approach the delta function centered on the distribution mean. 

The volume of the local area can be defined by sampling for each element from a gamma distribution
\begin{eqnarray}
    X_i \sim \Gamma(\alpha=\frac{n}{N},\beta=1)
\end{eqnarray}
where $\alpha$ and $\beta$ are the $\Gamma$ shape and rate parameters, $n$ is the number of atoms in the local environment, and $N$ is the number of distinct elements on the lattice site.

In this work, the depth of the local environment was determined from the fitting in Section~\ref{sec:sfe}. Sampling each (111) plane separately, the distribution of interface cation arrangements is built from $X_i\sim\Gamma(1,1)$, equating to a naive assumption of $n=N$. Without knowing the exact shape or size of the critical area, this assumption corresponds with a cutoff distance on the order of $2b$.

\begin{figure}[!t]
\centering
\includegraphics[width=.45\textwidth]{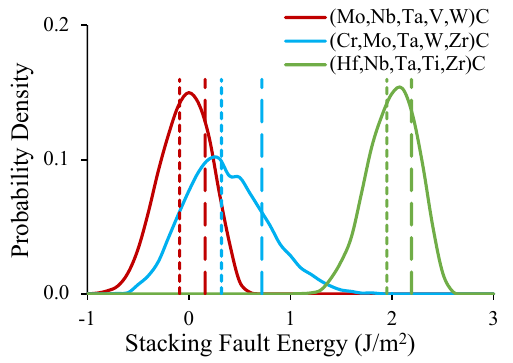}
\caption{\label{fig:dist} Predicted ISF distributions for three HETMCs. Dashed vertical lines represent the mean and the mean plus one standard deviation for each distribution.}
\end{figure}

The predicted ISF distributions derived from this $\Gamma$ sampling and the modified RoM model in Section~\ref{sec:sfe} are plotted in Fig.~\ref{fig:dist} for three different compositions along with the mean and mean plus one standard deviation for each distribution. For (Mo,Nb,Ta,V,W)C, the mean stacking fault energy and standard deviation are -0.09 J/m$^2$ and 0.25 J/m$^2$, respectively. Using $G=193$ GPa and $b=0.437$ nm~\cite{liu_stability_2022} in Eq.~\ref{eqn:partial} yields a Shockley partial separation $s$ of less than 18 nm. Hence, although 64\% of the stacking fault distribution is negative, the naive model predicts a finite partial separation.

\begin{figure}[!t]
\centering
\includegraphics[width=.45\textwidth]{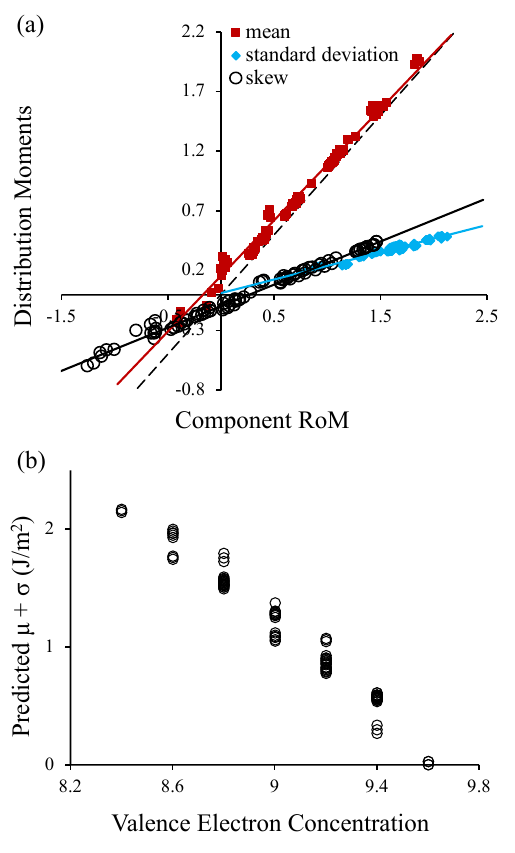}
\caption{\label{fig:moments} Mean, standard deviation, and skew (a) of the 126 predicted HETMC ISF energy distributions and (b) the sum of first two moments as a function of VEC.}
\end{figure}

The sampling procedure was carried out on each of the 126 HETMC compositions consisting of five-cation combinations from Cr, Hf, Mo, Nb, Ta, Ti, V, W, and Zr. The mean, standard deviation, and skew of the predicted ISF distributions are plotted in Fig.~\ref{fig:moments} as a function of the same quantities calculated from the five constituent single-metal carbides. There are strong linear relations for each of these distribution parameters, adding another facet to the stacking fault RoM. The linear coefficients and fitting errors are given in Table~\ref{tab:moments}.

\begin{table}
    \caption{\label{tab:moments} RoM linear regression coefficients and errors (J/m$^2$) for the first three moments of the predicted ISF distributions.}
    \centering
    \begin{ruledtabular}
    \begin{tabular}{lccc}
        & $\hat\beta_1$ & $\hat\beta_0$ & RMSE \\
        \hline
       Mean & 0.932 & 0.156 & 0.058 \\
       Standard deviation & 0.228 & 0.015 & 0.015 \\
       Skew & 0.361 & 0.096 & 0.026 \\
    \end{tabular}
    \end{ruledtabular}
\end{table}

Finally, the $\mu+\sigma$ stacking fault energies, as calculated from the predicted distribution moments, are plotted as a function of VEC in Fig.~\ref{fig:moments}b. There is a monotonic decrease in energy, suggesting that the separation between Shockley partial dislocations increases with increasing VEC. However, with the ``mean plus one'' model as a lower bound estimate of the energy barrier for dislocation pinning, partial dislocation separation distances are expected to be finite for all VEC values $\leq9.6$.

\section{Conclusion}

First principles calculations were used to establish new RoMs for interfacial defects in group IVB, VB, and VIB stoichiometric HETMCs. \{111\} ISF energies were shown to be predictable from an optimized RoM based on values from the respective single-metal carbide compositions. A linear relationship was discovered between these stacking fault energies and the USF energies along the $\langle11\bar2\rangle\{111\}$ slip path on the gamma surface. Provided that the shift of the USF barrier from the high-symmetry position is accounted for, the linear coefficient is independent of composition across the range of early transition metal carbides. The ISF and USF energies are analogous to the heat of reaction and transition state barrier in chemical reactions, and this linear relationship is similarly analogous to the EPS relationship that is ubiquitous across chemical systems, but not applied to plasticity in solids. 

It was further found that the mean, standard deviation, and skew for the distribution of stacking fault energies resulting from cation disorder around the boundary are predictable from linear relations that connect the stacking fault energies of the single-metal carbides to the corresponding high-entropy compositions. Using a model where the upper tail of the ISF distribution balances the elastic repulsion between partial dislocations, we demonstrate that Shockley partial edge dislocations should remain bound for HETMC compositions with \mbox{VEC $\leq9.6$} despite certain compositions exhibiting negative mean ISF energies. This is consistent with experiment, where compositions predicted to have negative stacking fault energies, such as (Mo,Nb,Ta,V,W)C, are phase stable with relatively low defect densities~\cite{sarker_high-entropy_2018}.

The RoMs and linear relations discovered in this study are important for understanding and predicting the defect structure and mechanical properties of HETMCs. In addition, the EPS relation has been an invaluable tool for estimating reaction kinetics across diverse areas of chemistry, and based on our results we anticipate that it will find other applications in predicting mechanical properties of high-entropy materials. Future work on this topic should include vacancy effects on gamma surface energies, and may require more sophisticated machine learning techniques for quantitative predictability~\cite{zhao_machine_2023, nam_prediction_2023}.

\begin{acknowledgments}

The authors gratefully acknowledge the Office of Naval Research for financial support of this work through the DoD SPICES MURI (Naval Research contract N00014-21-1-2515) and Naval Research grant N00014-23-1-2758. We also acknowledge the computing resources provided by North Carolina State University High Performance Computing Services Core Facility (RRID:SCR\_022168).

\end{acknowledgments}


\providecommand{\noopsort}[1]{}\providecommand{\singleletter}[1]{#1}%

\end{document}